\documentclass[conference]{IEEEtran}
\IEEEoverridecommandlockouts
\usepackage[frozencache,cachedir=.]{minted} %
\usepackage{cite}
\usepackage{amsmath,amssymb,amsfonts,amsthm}
\usepackage{algorithmic}
\usepackage{graphicx}
\usepackage{textcomp}
\usepackage{xcolor}
\usepackage{braket}
\usepackage{bbm}
\usepackage{minted}
\usepackage{caption}
\usepackage{subcaption}
\def\BibTeX{{\rm B\kern-.05em{\sc i\kern-.025em b}\kern-.08em
    T\kern-.1667em\lower.7ex\hbox{E}\kern-.125emX}}
    
\usepackage[breaklinks=true, colorlinks=true, linkcolor={blue!90!black}, citecolor={blue!90!black}, urlcolor={blue!90!black}]{hyperref}
\usepackage{cite,doi,blindtext}

\newtheorem*{portfolio_optimization}{Portfolio Optimization}
\newtheorem*{vertex_cover}{Minimum Vertex Cover}
\newtheorem*{clique}{Maximum Clique}
\newtheorem*{max_bisection}{Maximum bisection}
\newtheorem*{graph_partition}{Graph Partition}

\usepackage{xspace}
\newcommand{\catchyname}{QVoice\xspace}

\newcommand{\bx}{\mathbf{x}}
\newcommand{\B}{\mathbb{B}}

\begin{document}

\title{Exploiting In-Constraint Energy in Constrained Variational Quantum Optimization}

\author{\IEEEauthorblockN{Tianyi Hao\IEEEauthorrefmark{1}\IEEEauthorrefmark{2}, Ruslan Shaydulin\IEEEauthorrefmark{3}, Marco Pistoia\IEEEauthorrefmark{3}, Jeffrey Larson\IEEEauthorrefmark{1}}\vspace{0.05in}
\IEEEauthorblockA{\IEEEauthorrefmark{1}
Argonne National Laboratory, Lemont, IL 60439 USA}
\IEEEauthorblockA{\IEEEauthorrefmark{2}
University of Wisconsin-Madison, Madison, WI 53706 USA}
\IEEEauthorblockA{\IEEEauthorrefmark{3}
JPMorgan Chase, New York, NY 10017 USA}
}

\maketitle

\begin{abstract}
A central challenge of applying near-term quantum optimization algorithms to industrially relevant problems is the need to incorporate complex constraints. In general, such constraints cannot be easily encoded in the circuit, and the quantum circuit measurement outcomes are not guaranteed to respect the constraints. Therefore, the optimization must trade off the in-constraint probability and the quality of the in-constraint solution by adding a penalty for constraint violation into the objective. We propose a new approach for solving constrained optimization problems with unconstrained, easy-to-implement quantum ans{\"a}tze. Our method leverages the in-constraint energy as the objective and adds a lower-bound constraint on the in-constraint probability to the optimizer. We demonstrate significant gains in solution quality over directly optimizing the penalized energy. We implement our method in \catchyname, a Python package that interfaces with Qiskit for quick prototyping in simulators and on quantum hardware.
\end{abstract}

\section{Introduction}\label{s1}

Combinatorial optimization abstracts many real-world optimization problems and has crucial applications in various fields. A combinatorial optimization problem is commonly specified by an objective function $f$ defined on the Boolean cube $\B$ and a set of feasible solutions $F\subseteq \B$. The goal is to find a solution $\bx\in F$ giving the maximum objective function value over all feasible solutions $\max_{\bx \in F} f(\bx)$. The set of feasible solutions is commonly described by one or more constraints. In most cases, $F$ is too large to explore exhaustively, and many combinatorial optimization problems are hard to solve classically.

Quantum computing provides a novel computational paradigm for solving combinatorial optimization problems. The implementation of quantum algorithms, however, is limited by small-scale, noisy, and error-prone contemporary hardware. Variational quantum algorithms (VQAs) such as the variational implementation of the quantum approximate optimization algorithm (QAOA)~\cite{Hogg2000,farhi2014quantum} and the variational quantum eigensolver (VQE)~\cite{cerezoVariationalQuantumAlgorithms2021} are promising algorithms for solving optimization problems on near-term quantum computers because of their relatively low resource requirements. These algorithms combine a parameterized quantum evolution with a classical optimization routine that finds parameters such that the measurement outcomes of the quantum evolution correspond to high-quality solutions to the optimization problem with high probability. %

Solving combinatorial optimization problems with VQAs requires choosing a technique for optimizing the quantum evolution parameters. In the cases where the measurement outcomes of the quantum evolution are guaranteed to satisfy the constraints (e.g., when the constraints are enforced by the quantum circuit), directly optimizing the expected value of the objective $f$ is sufficient. However, enforcing the constraints in the circuit in many cases leads to high circuit depth~\cite{hadfield2018quantum,Herman2022Zeno}, making the resulting circuits hard to execute on near-term noisy hardware~\cite{Niroula2022}. Therefore, an alternative approach is commonly used, wherein the circuit is not required to respect the constraints and a penalty is added to the objective to drive the parameter optimization to the in-constraint subspace. If the problem has only one constraint, the objective becomes $\max_{\bx\in\mathbb{B}}f(\bx)+\lambda g(\bx)$, where $g(\bx)$ is the function encoding the constraint (e.g., $g(\bx) = 0$ if $\bx\in F$, and $1$ otherwise) and $\lambda$ is the coefficient controlling the strength of the penalty. This approach generalizes trivially to multiple constraints by adding multiple penalties.
Choosing the values for $\lambda$ is extremely important in most problems and is generally not an easy task. $\lambda$ being too large often leads to a nearly uniform mixture of feasible states, as the optimizer focuses on penalizing out-constraint states; $\lambda$ being too small tends to result in a state with low in-constraint probability, where most solution samples are infeasible. The necessity of choosing a good value for the penalty coefficient introduces high computational overhead, since in general the value must be tuned independently for each problem instance.

In this work we propose a novel approach for solving constrained optimization problems with VQAs using ans\"atze that do not preserve constraints. Instead of enforcing the constraints by adding a penalty, we optimize the original objective $f$ with only in-constraint samples and an optimizer constraint on the minimum value of the in-constraint probability. This optimizer constraint guarantees a lower bound on the fraction of sampling feasible solutions. We observe significantly improved performance compared with the penalty method, without the overhead of expensive parameter tuning. We implement our method as an open-source Python package \catchyname that integrates with IBM Qiskit, available at \url{https://github.com/HaoTy/QVoice}.

The rest of the paper is organized as follows. Section~\ref{s2} reviews the key concepts of VQE and QAOA and existing approaches for using them to solve constrained optimization problems. We introduce the notion of in-constraint energy, the details of our method, and \catchyname in Section~\ref{s3}. Section~\ref{s4} presents and analyzes the results from numerical experiments. In Section~\ref{s5} we conclude with a discussion of the importance of this work and the opportunities for quantum advantage in combinatorial optimization.

\begin{figure}[t]
    \centering
    \includegraphics[width=0.4\textwidth]{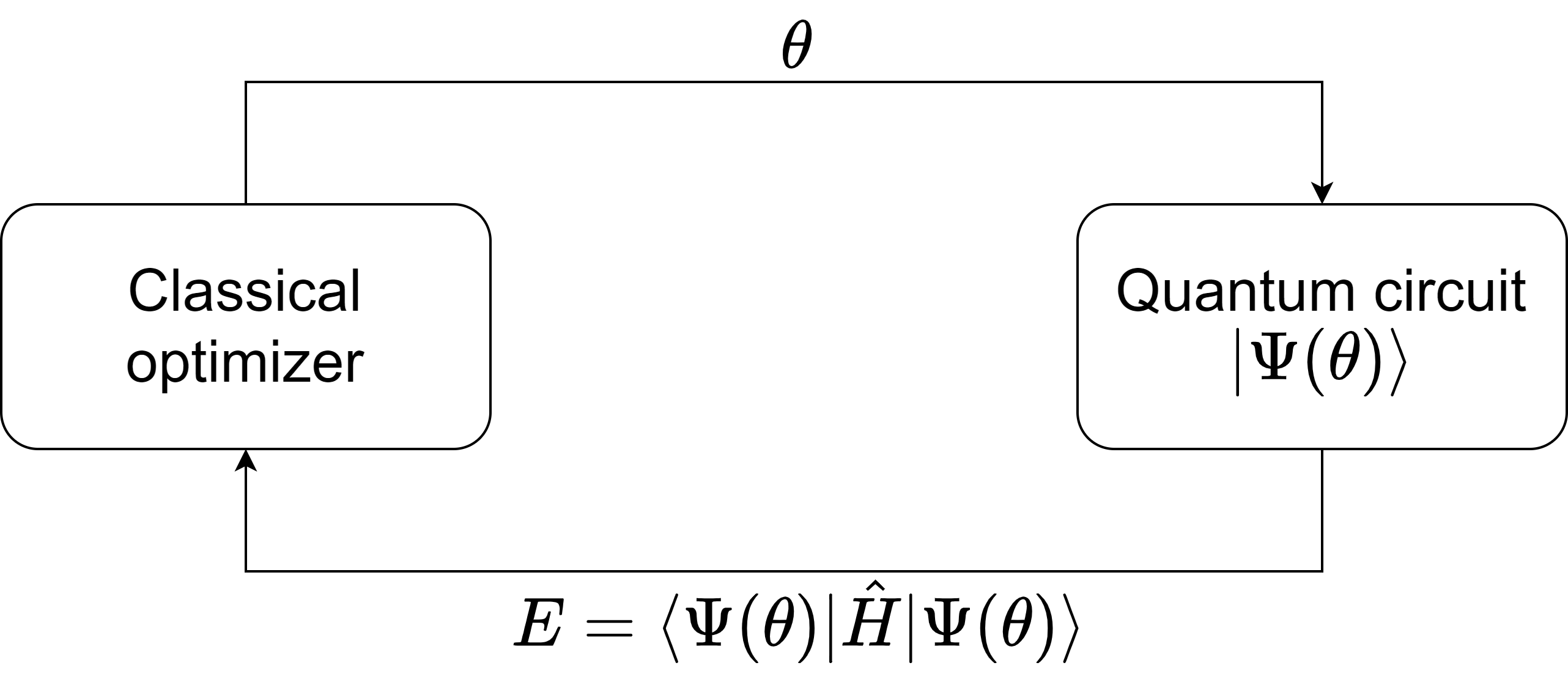}
    \caption{Workflow of variational quantum algorithms. The parameterized quantum circuit outputs the energy $E=\braket{\Psi(\theta)|\hat{H}|\Psi(\theta)}$, which is used by the classical optimizer to update the parameter $\theta$.}
    \label{fig:workflow}
\end{figure}

\section{Background}\label{s2}

The variational quantum algorithms (VQAs) we consider solve optimization problems by combining a parameterized quantum circuit $\ket{\Psi(\theta)}$ (``ans\"atz'') with a classical optimization routine to find $\theta$, such that the measurement outcomes of $\ket{\Psi(\theta)}$ correspond to good solutions of the original problem with high probability. To realize the mapping between the problem solution space and the Hilbert space, one can construct a Hamiltonian $\hat{H}$ by substituting the binary variables in the problem's objective and constraint functions with $(\hat{Z}-\hat{I})/2$, where $\hat{Z}$ is the Pauli Z operator and $\hat{I}$ is the identity operator. The algorithm then seeks to find the ground state of $\hat{H}$ by going through several optimization iterations.
For each iteration, the quantum circuit is executed multiple times to obtain a number of samples of the expectation value of the Hamiltonian $\braket{\Psi(\theta)|\hat{H}|\Psi(\theta)}$, also known as the energy of the state $\ket{\Psi(\theta)}$. The classical optimizer uses the sample mean of the energy as its objective and outputs the updated parameter $\theta$ to the quantum circuit. This process is illustrated in Figure \ref{fig:workflow}.

The variational quantum eigensolver (VQE) was the first VQA proposed, originally designed for finding the ground state of a given molecule \cite{peruzzo2014variational}. It has subsequently been extended with various ans\"atze to solve a broad range of problems \cite{cerezoVariationalQuantumAlgorithms2021}. The two most popular groups of ans\"atze
are the chemistry-inspired ans\"atze and the hardware-efficient ans\"atze \cite{fedorovVQEMethodShort2022, kandala2017hardware,Liu2022}.
In this work we use the problem-independent hardware-efficient Two Local ans{\"a}tz with $R_y$ rotation blocks, $CZ$ entanglement blocks, and one linear entanglement layer.

The quantum approximate optimization algorithm (QAOA) is a hybrid quantum-classical algorithm for solving combinatorial optimization problems \cite{farhi2014quantum}, where the ans\"atz is inspired by the adiabatic evolution \cite{farhi2000quantum, farhi2001quantum}. The ans\"atz is constructed by applying pairs of alternating operators to a uniform superposition over all computational basis states as
\begin{equation}
    \ket{\text{QAOA}} = \prod_{j=1}^pe^{-i\beta \hat{B}}e^{-i\gamma \hat{C}}\ket{+}^{\otimes N},\label{qaoa}
\end{equation}
where $\hat{C} = diag(f(x))$ encodes the objective and $\hat{B}=\sum_{j}X_j$ is the sum of single-qubit Pauli $\hat{X}$. 

Various strategies for identifying high-quality VQE and QAOA parameters have been proposed in the past, including the use of reinforcement learning~\cite{khairy2019learning,Wauters2020} and alternative objective functions~\cite{Barkoutsos2020, kolotourosEvolvingObjectiveFunction2021}. While many of these strategies can be adapted to handle constraints, to the best of our knowledge no parameter optimization strategy has considered the need to enforce constraints explicitly.

\begin{figure}[t]
    \centering
    \includegraphics[width=0.4\textwidth]{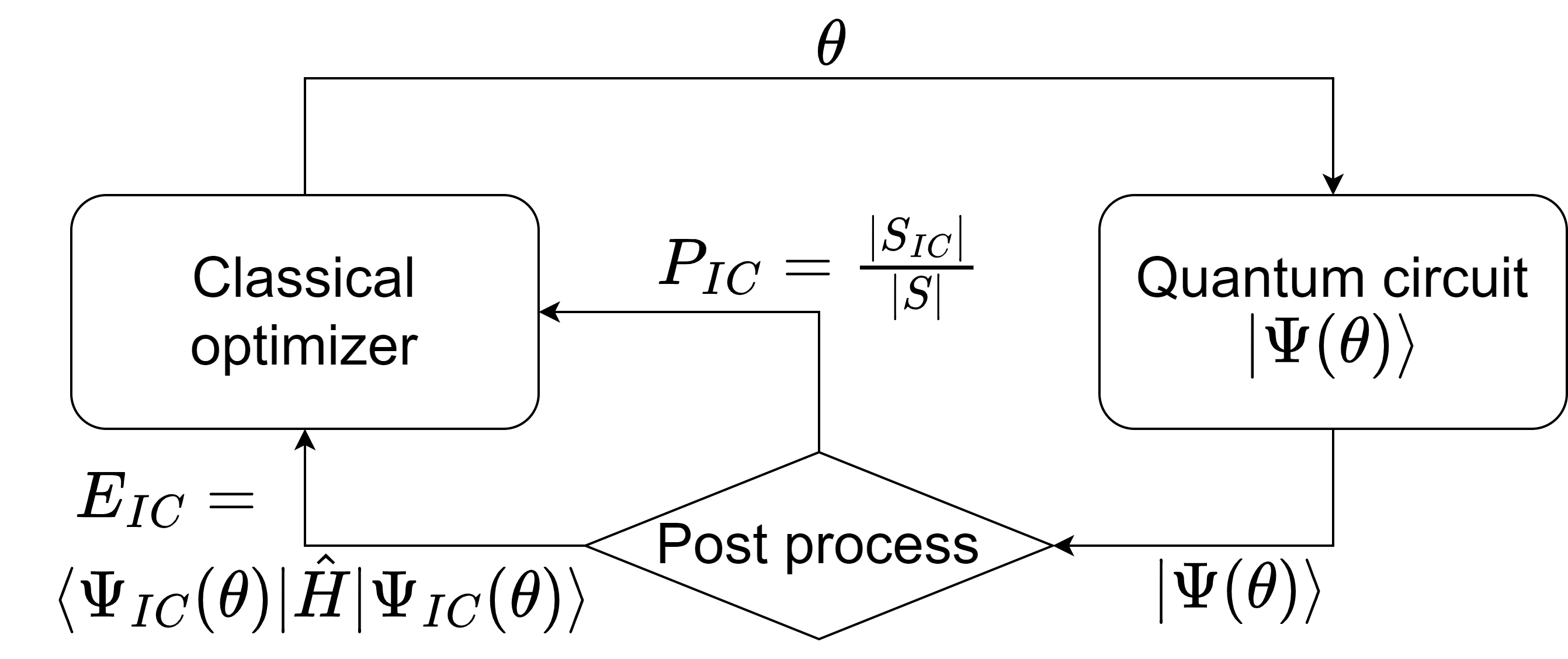}
    \caption{Workflow of variational quantum algorithms using in-constraint energy $E_{IC}=\braket{\Psi_{IC}(\theta)|\hat{H}|\Psi_{IC}(\theta)}$ as the objective and having in-constraint probability $P_{IC}$ as additional information for the optimizer.}
    \label{fig:workflow2}
\end{figure}

\section{\catchyname: Quantum Variational Optimization with In-Constraint Energy}\label{s3}

In variational quantum algorithms, using the expectation of the objective function (equivalently, the energy of the Hamiltonian encoding the objective) as the objective for parameter optimization works well for unconstrained problems.
For constrained problems, however, if the constraints are also encoded into the Hamiltonian in the form of penalty terms, then optimizing the energy does not truthfully reflect the original problem's objective. To faithfully measure the quality of solutions for constrained problems, we employ the notions of approximation ratio and in-constraint probability as metrics. Given a collection of solution samples $S$ and an in-constraint subcollection $S_{IC}=S\cap F$, the in-constraint probability 
\begin{align}
    P_{IC} = \frac{|S_{IC}|}{|S|}
\end{align}
is the proportion of feasible samples in all samples. Assuming the problem's objective $f$ needs to be minimized, we can define the approximation ratio as 
\begin{align}
    \rho = \frac{f_{\max} - \sum_{\mathbf{s}\in S_{IC}} f(\mathbf{s}) / |S_{IC}|}{f_{\max} - f_{\min}},
\end{align}
where $f_{\max}$ and $f_{\min}$ are, respectively, the maximum and minimum objective function values achievable without violating any constraints.

\begin{figure*}[t]
    \centering
    \captionsetup[subfigure]{width=0.85\linewidth}
    \begin{subfigure}[t]{0.329\linewidth}
        \centering
        \includegraphics[width=\linewidth]{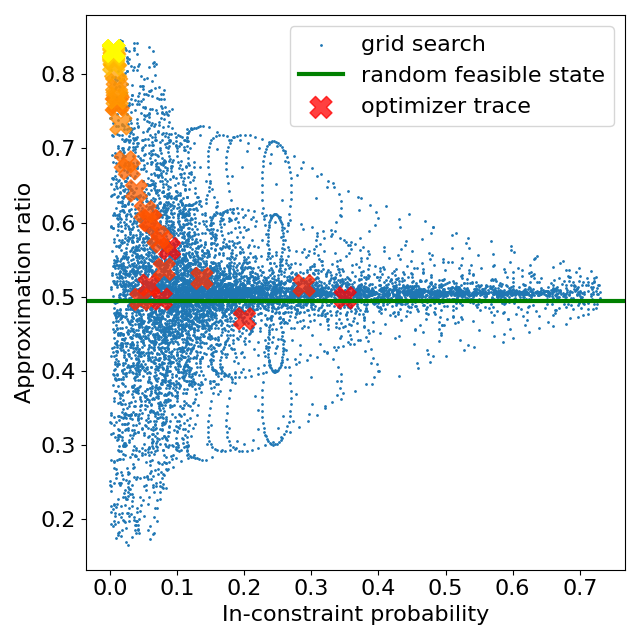}
        \caption{Using in-constraint energy as the objective.}
    \end{subfigure}
    \begin{subfigure}[t]{0.329\linewidth}
        \centering
        \includegraphics[width=\linewidth]{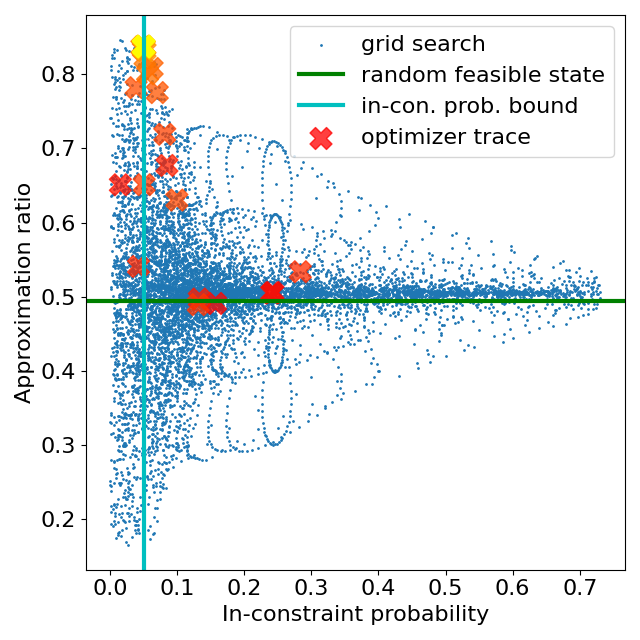}
        \caption{Using in-constraint energy as the objective with a lower-bound constraint on in-constraint probability.}
    \end{subfigure}
    \begin{subfigure}[t]{0.329\linewidth}
        \centering
        \includegraphics[width=\linewidth]{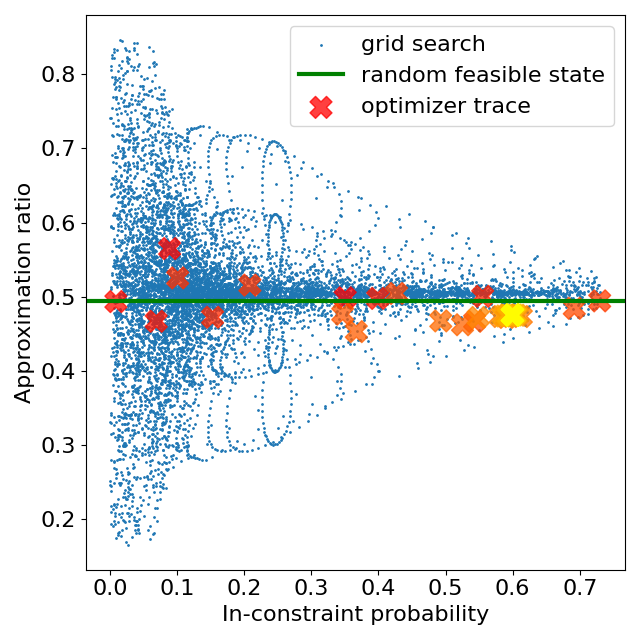}
        \caption{Using energy as the objective with penalty terms in the Hamiltonian.}
    \end{subfigure}
    \caption{Data points from QAOA grid search plotted as approximation ratio versus in-constraint probability, overlaid with optimizer trace, for different objective settings: (a) in-constraint energy, (b) in-constraint energy with lower-bound constraint on in-constraint probability, and (c) energy. Color of the trace varies from the beginning of the optimization (red) to the end (yellow). }
    \label{fig:grid_search}
\end{figure*}

The noticeable difference between the energy and the approximation ratio is that the latter takes into account only the feasible solutions, which makes it a better metric in the constrained optimization setting. We cannot directly use it as the objective since we do not know the $f_{\max}$ and $f_{\min}$ without solving the problem. Nevertheless, having only the nonconstant part $\sum_{\mathbf{s}\in S_{IC}} f(\mathbf{s}) / |S_{IC}|$ serves the same purpose for the optimizer.
Following this reasoning, we introduce the in-constraint energy 
\begin{align}
    E_{IC}=\braket{\Psi_{IC}|\hat{H}|\Psi_{IC}}
\end{align}
as an alternative objective for constrained problems, where given a sampled state $\ket{\Psi} = \sum_{\mathbf{s}}c_{\mathbf{s}}\ket{s}$, 
\begin{align}
    \ket{\Psi_{IC}} = \frac{\sum_{\mathbf{s}}\mathbbm{1}_F(\mathbf{s})c_{\mathbf{s}}\ket{s}}{|\sum_{\mathbf{s}}\mathbbm{1}_F(\mathbf{s})c_{\mathbf{s}}\ket{s}|}
\end{align} 
is the normalized in-constraint state. 
If the full amplitude information is available, for example with a state vector simulator, it can be implemented by simply removing the amplitudes of the infeasible bases and normalizing the remaining state. When implementing it on a quantum device, we can calculate the expectation value with the post processed samples either by looping over the Pauli terms in the Hamiltonian or by preparing a state that approximates the samples. 

We note that using the in-constraint energy as the objective incurs an overhead for processing the samples. In order to be computationally tractable, our method requires that, given a solution, the problem constraint functions can be computed in polynomial time with respect to the number of qubits $n$ and that the number of samples is a polynomial in $n$. The former is true for most constrained combinatorial optimization problems, and the latter is a common practice and a reasonable assumption. One can also store the computed feasibility information to trade space for time, which can almost remove the overhead after some iterations, when the sampled states are concentrated.

Since the in-constraint energy contains no information about the infeasible states, the optimizer does not implicitly keep the in-constraint probability high as it does with the penalty method. This situation can be solved by explicitly adding a lower-bound constraint on the in-constraint probability to the optimizer. The full workflow is shown in Figure~\ref{fig:workflow2}.

Figure~\ref{fig:grid_search} shows a single $10$-variable portfolio optimization instance solved by the $1$-layer QAOA with the default mixer as in \eqref{qaoa} and $300$-iteration COBYLA as the optimizer. The data points of the grid search are plotted in a multi-objective optimization fashion, where the $x$-axis is the in-constraint probability and the $y$-axis is the approximation ratio, in combination with an overlay of the optimizer trace of three objective settings: in-constraint energy, in-constraint energy with lower-bound constraint on in-constraint probability, and energy with Qiskit's default penalty heuristic. We observe that although QAOA with in-constraint energy achieves an approximation ratio significantly higher than the baseline, the final in-constraint probability is too small to sample enough feasible solutions. By explicitly setting a lower-bound constraint for the in-constraint probability, we can balance between the two objectives and reach any desired position on the Pareto frontier. The default QAOA suffers from the need to satisfy the penalty term and thus ends up with a high in-constraint probability and a near-baseline approximation ratio.

We implement the in-constraint energy objective and the lower-bound constraint on the in-constraint probability as a Python package \catchyname that extends the variational quantum optimization functionalities of Qiskit. Users can define the problem and the variational algorithm with Qiskit as usual, fully utilize what Qiskit has to offer, and change the objective to in-constraint energy with one line of code.
\begin{minted}[breaklines,fontsize=\scriptsize,baselinestretch=1.2]{python}
from qvoice import InConstraintSolver
from qiskit_aer import AerSimulator
from qiskit.algorithms import QAOA
from qiskit.algorithms.optimizers import COBYLA
from qiskit_finance.applications.optimization import PortfolioOptimization

problem = PortfolioOptimization(...).to_quadratic_program()
algorithm = QAOA(optimizer=COBYLA(), quantum_instance=AerSimulator(method="statevector"))

result = InConstraintSolver(algorithm, problem).solve()
\end{minted}
Beyond the minimalistic usage example, \catchyname provides options for customization, logging, and plotting, as well as integration with other optimization libraries.
\catchyname is open-source, available at \url{https://github.com/HaoTy/QVoice}. 

\section{Performance Improvements}\label{s4}
\begin{figure*}[t]
    \centering
    \begin{subfigure}[t]{0.49\linewidth}
        \centering
        \includegraphics[width=\linewidth]{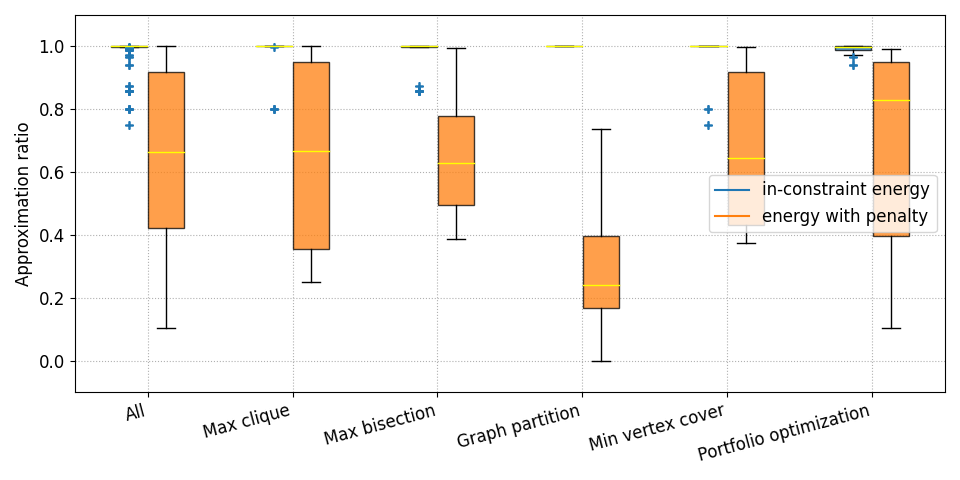}
        \includegraphics[width=\linewidth]{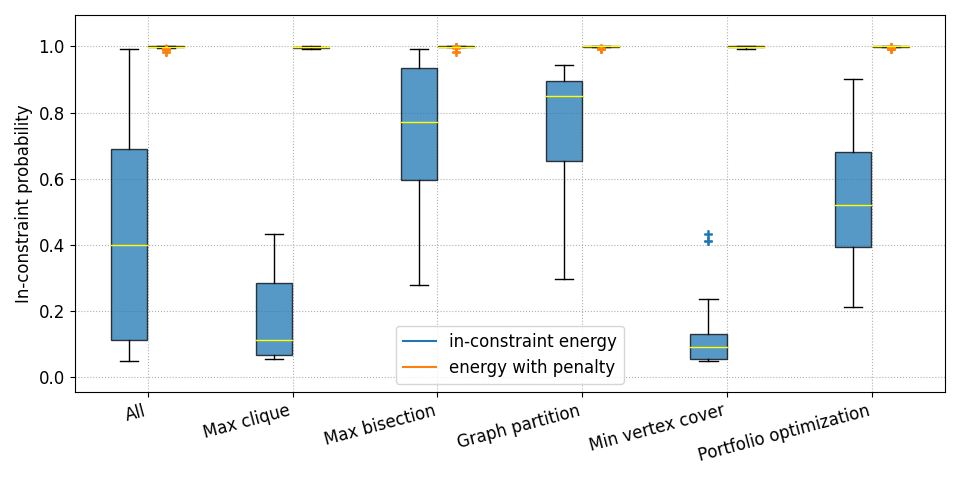}
        \includegraphics[width=\linewidth]{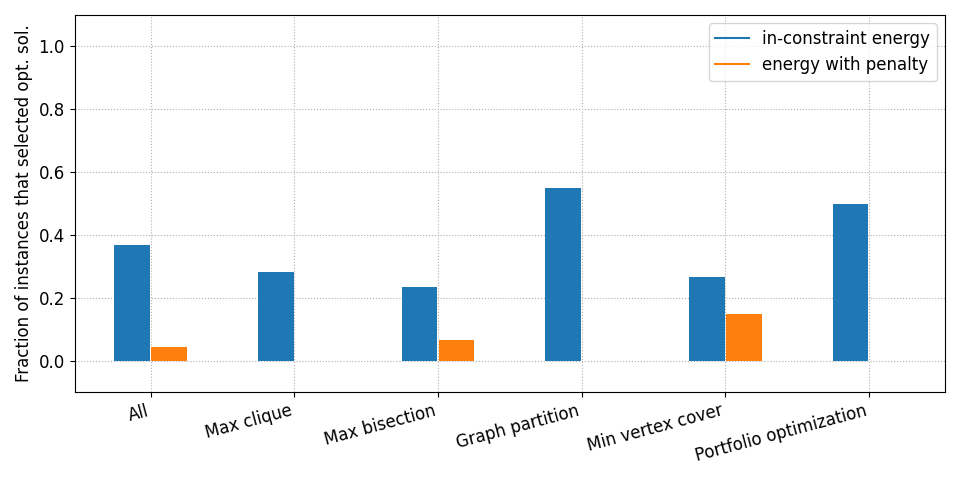}
        \caption{VQE}
    \end{subfigure}
    \begin{subfigure}[t]{0.49\linewidth}
        \centering
        \includegraphics[width=\linewidth]{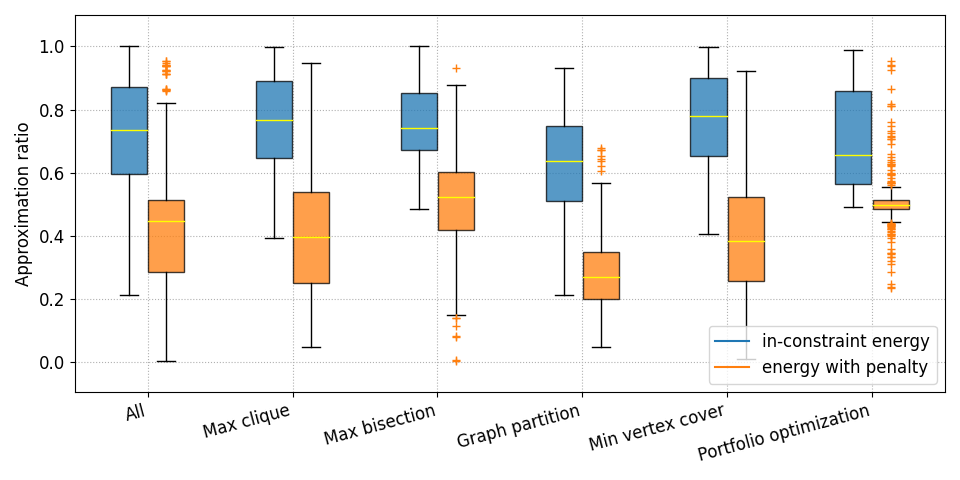}
        \includegraphics[width=\linewidth]{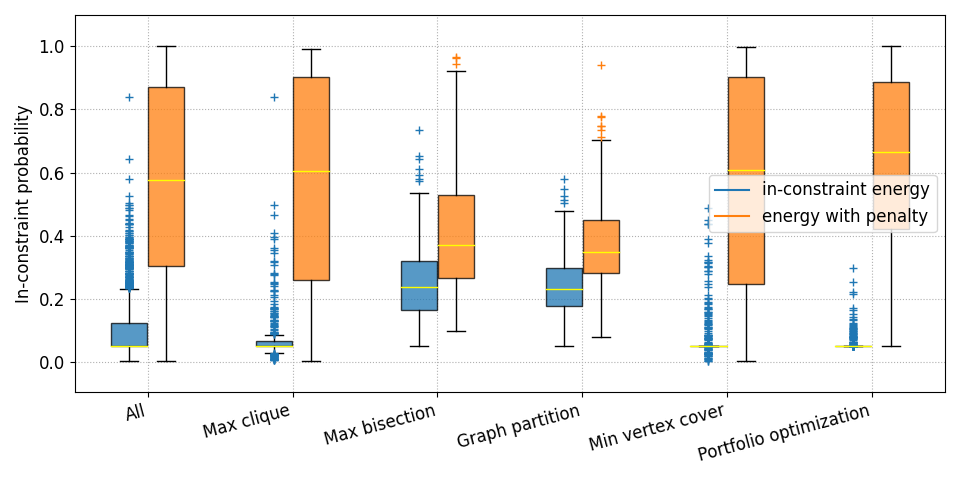}
        \includegraphics[width=\linewidth]{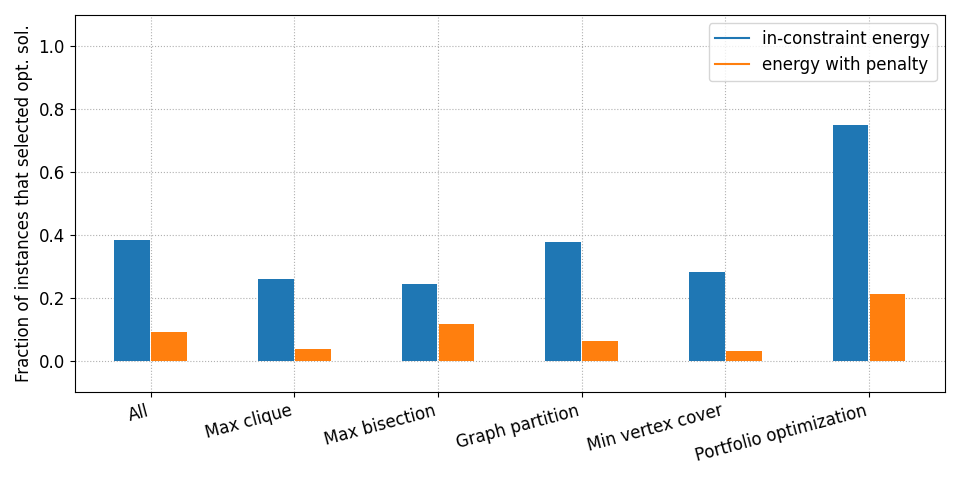}
        \caption{QAOA}
    \end{subfigure}
    \caption{Comparison of the solution quality between using energy with penalty and in-constraint energy as the objective function for (a) VQE and (b) QAOA. Three metrics are plotted: box plots of the approximation ratio (top) and in-constraint probability (middle) and the bar plot of the fraction of instances that have the optimal solution as the most sampled feasible solution (bottom). Our approach leads to higher solution quality while still producing adequate feasible samples.}
    \label{fig:box}
\end{figure*}

\begin{figure*}[ht]
    \centering
    \begin{subfigure}[t]{\linewidth}
    \centering
    \includegraphics[width=0.329\linewidth]{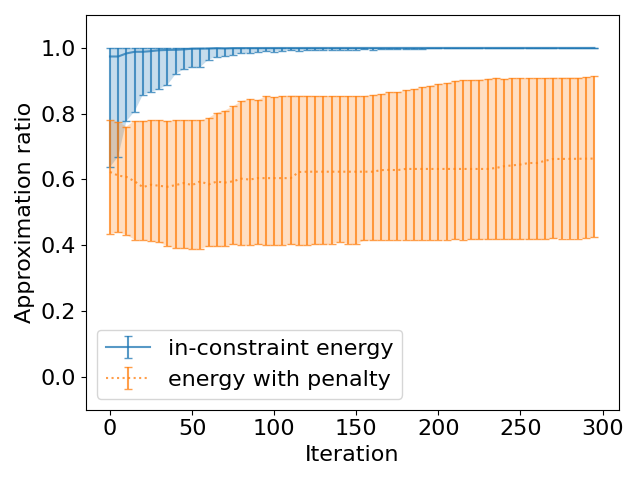}
    \includegraphics[width=0.329\linewidth]{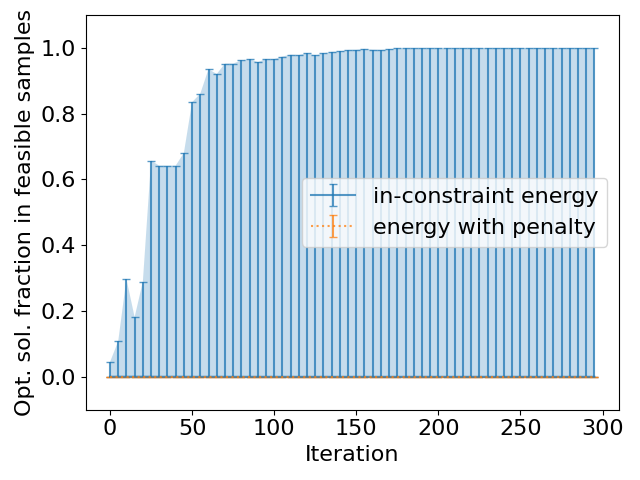}
    \includegraphics[width=0.329\linewidth]{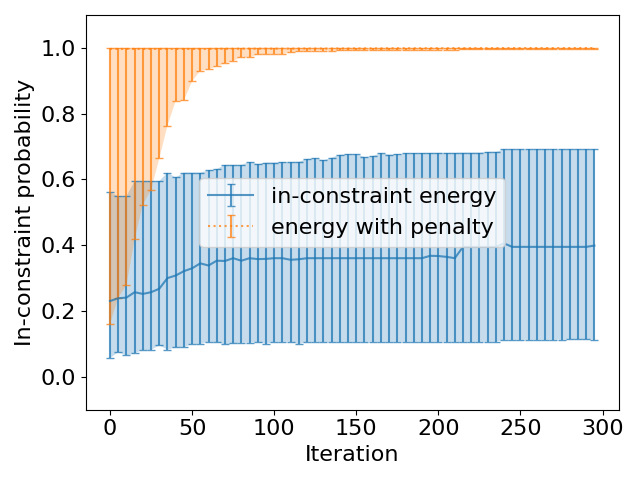}
    \caption{VQE}
    \end{subfigure}
    \begin{subfigure}[t]{\linewidth}
    \centering
    \includegraphics[width=0.329\linewidth]{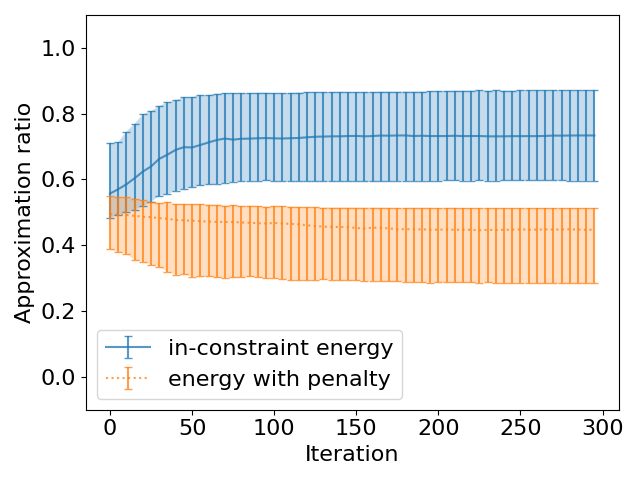}
    \includegraphics[width=0.329\linewidth]{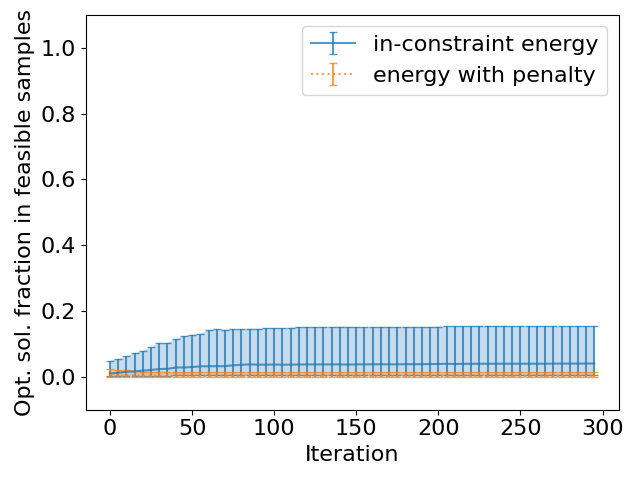}
    \includegraphics[width=0.329\linewidth]{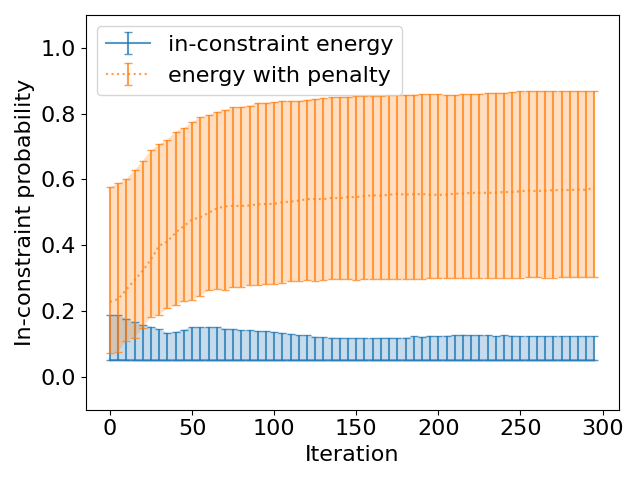}
    \caption{QAOA}
    \end{subfigure}
    \caption{Comparison of the optimization process between using energy with penalty and in-constraint energy as the objective function for (a) VQE and (b) QAOA. All problem instances are aggregated to plot quartiles of three metrics versus the number of iterations: approximation ratio (left), fraction of optimal solution in feasible samples (middle), and in-constraint probability (right).}
    \label{fig:error bars}
\end{figure*}

\begin{figure*}[ht]
    \centering
    \includegraphics[width=0.329\linewidth]{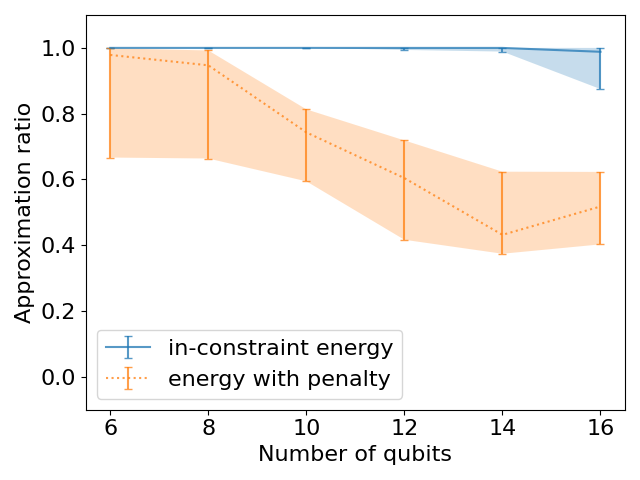}
    \includegraphics[width=0.329\linewidth]{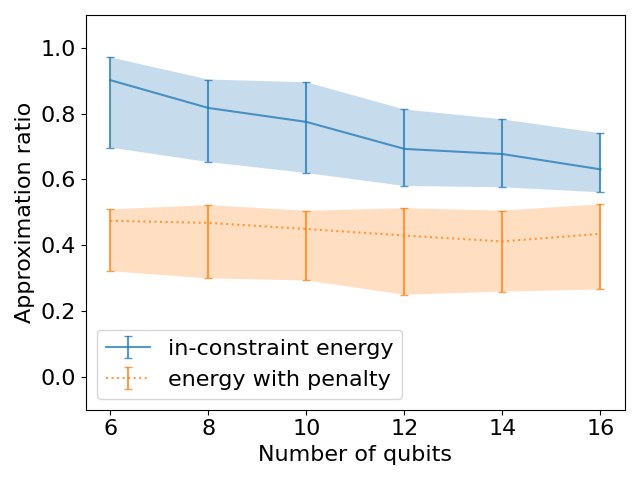}
    \includegraphics[width=0.329\linewidth]{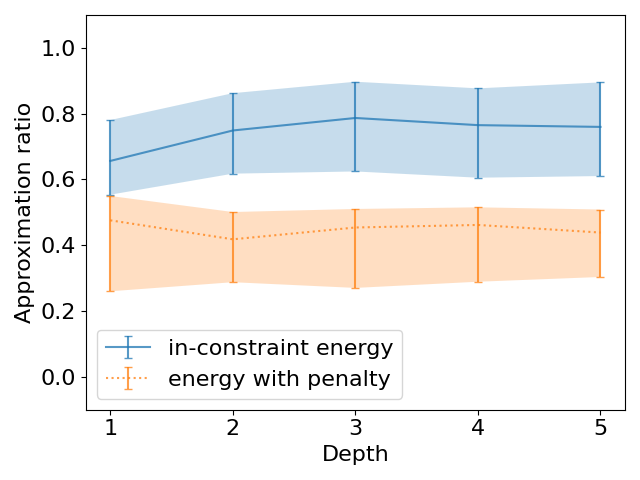}
    \caption{Quartile plots of the final approximation ratio of VQE with different numbers of qubits (left), QAOA with different numbers of qubits (middle), and QAOA with different depths (right). All problem instances with the same number of qubits or depths are aggregated.}
    \label{fig:vary}
\end{figure*}

We test our method empirically with five constrained combinatorial optimization problems that are of wide interest: maximum clique, minimum vertex cover, maximum bisection, graph partition, and portfolio optimization. 
They can be formally defined as follows:
\begin{clique}
Given an undirected graph $G=(V,E)$ with weight $w_i$ on vertex $v_i$, find a subset $V'$ of $V$ such that every pair of vertices in $V'$ is connected by an edge in $E$ and the sum of weights of vertices in $V'$ is maximized.
Equivalently, the problem is to find $\mathbf{x} \in \{0,1\}^{|V|}$ that will
\begin{align*}
    &\text{\upshape maximize }&&\mathbf{w}^T \mathbf{x}\\
    &\text{\upshape subject to: }&&x_i + x_j \le 1\text{ for }i, j\in [|V|], i\neq j, (v_i,v_j)\not\in E.
\end{align*}
\end{clique}
\begin{vertex_cover}
Given an undirected graph $G=(V,E)$ with weight $w_i$ on vertex $v_i$, find a subset $V'$ of $V$ such that every edge in $E$ has at least one endpoint in $V'$ and the sum of weights of vertices in $V'$ is minimized.
Equivalently, the problem is to find $\mathbf{x} \in \{0,1\}^{|V|}$ that will
\begin{align*}
    &\text{\upshape minimize }&&\mathbf{w}^T \mathbf{x}\\
    &\text{\upshape subject to: }&&x_i + x_j \ge 1\text{ for }i, j\in [|V|], (v_i,v_j)\in E.
\end{align*}
\end{vertex_cover}

\begin{max_bisection}
Given an undirected graph $G=(V,E)$ with weight $w_e$ on edge $e$, find two complementary subsets $V_1$ and $V_2$ of $V$ such that $|V_1|=|V_2|=|V|/2$ (assuming $|V|$ is even) and the sum of weights of edges between $V_1$ and $V_2$ is maximized.
Equivalently, the problem is to find $\mathbf{x} \in \{0,1\}^{|V|}$ that will
\begin{align*}
    &\text{\upshape maximize }&&w_{(v_i,v_j)} x_i x_j\text{ for }i, j\in [|V|], (v_i,v_j)\in E\\
    &\text{\upshape subject to: }&&\mathbf{1}^T \mathbf{x} = |V|/2.
\end{align*}
\end{max_bisection}

\begin{graph_partition}
Given an undirected graph $G=(V,E)$ with weight $w_e$ on edge $e$, find two complementary subsets $V_1$ and $V_2$ of $V$ such that $|V_1|=|V_2|=|V|/2$ (assuming $|V|$ is even) and the sum of weights of edges between $V_1$ and $V_2$ is minimized.
Equivalently, the problem is to find $\mathbf{x} \in \{0,1\}^{|V|}$ that will
\begin{align*}
    &\text{\upshape minimize }&&w_{(v_i,v_j)} x_i x_j\text{ for }i, j\in [|V|], (v_i,v_j)\in E\\
    &\text{\upshape subject to: }&&\mathbf{1}^T \mathbf{x} = |V|/2.
\end{align*}
\end{graph_partition}

\begin{portfolio_optimization}
Given the number of assets $n$, expected returns $\mathbf{\mu} \in \mathbb{R}^n$, the covariance matrix of the returns $\mathbf{\Sigma} \in \mathbb{R}^{n\times n}$, a risk factor $q \in \mathbb{R}$, and a budget $B \in \mathbb{R}$, find $\mathbf{x} \in \{0,1\}^n$ that will
\begin{align*}
    &\text{\upshape minimize }&&q \mathbf{x}^T \mathbf{\Sigma} \mathbf{x} - \mathbf{\mu}^T \mathbf{x}\\
    &\text{\upshape subject to: }&&\mathbf{1}^T \mathbf{x} = B.
\end{align*}
\end{portfolio_optimization}

For the maximum clique problem and the minimum vertex cover problem, we use random $G(n,m)$ graphs with $m$ set to be $n(n-1)/4$, in other words, having half the edges of a complete graph. For the maximum bisection problem, we use random $3$-regular graphs, which are commonly used in its generalized unconstrained version, the maximum cut problem. For the graph partition problem, we use a social network model, the random planted 2-partition graph with in-group probability $1$ and out-group probability $2/n$, where $n$ is the number of vertices. This is equivalent to having two cliques with $n/2$ vertices each and probability $2/n$ to form an edge for each pair of vertices between them. 
The weights in all graphs are drawn from a normal distribution with mean $1$ and standard deviation $1e-4$ to create a unique optimal solution.
For the portfolio optimization problem, we use Qiskit to randomly generate mock stock-market data with $q=0.5$ and $B=n/2$.

We perform classically simulated numerical experiments with Qiskit~\cite{Qiskit}, using the state vector simulator. For both VQE and QAOA, we generate random instances of each problem with $6$, $8$, $10$, $12$, $14$, and $16$ variables and run the COBYLA optimizer for $300$ iterations using energy with Qiskit's default penalty heuristic and in-constraint energy as the objective, respectively. For the in-constraint energy approach, a lower bound constraint of $0.05$ is enforced on the in-constraint probability in the optimizer. We log the approximation ratio, in-constraint probability, and fraction of optimal solution in feasible samples of the intermediate states during the optimization process as metrics.
For VQE, we use the Two Local ans{\"a}tz with $R_y$ rotation blocks, $CZ$ entanglement blocks, and one linear entanglement layer. We generate $20$ instances of each problem under each setting. For QAOA, we construct the ans{\"a}tz with the penalized Hamiltonian for both energy and in-constraint energy to make the QAOA landscape consistent. We point out that ans{\"a}tze constructed with the penalized Hamiltonian seem to be generally more expressible than those constructed with the penalty-free Hamiltonian in our observation. We vary the QAOA depth from one to five and generate $10$ instances of each problem under each setting.

Figure \ref{fig:box} shows the box and bar plots of the metrics of the final states obtained by VQE and QAOA for each problem. We observe significant improvements in the quality of solutions for both algorithms. VQE with in-constraint energy can almost always achieve a nearly perfect approximation ratio, while the penalty method is very instance-dependent and generally does not obtain satisfying results. For QAOA, since the ans{\"a}tze are less expressive, both methods present overall worse performance. Nonetheless, QAOA with in-constraint energy still shows a clear advantage over the penalty method for all problems.

From the in-constraint probability plots, we see that the penalty method fails to find the optimal solution in most cases because of the distraction from the penalty term. For VQE, the maximum clique problem and the minimum vertex cover problem have more complicated constraints and a more confined feasible solution space compared with other problems, resulting in our method hitting the $0.05$ lower bound constraint on the in-constraint probability. For QAOA, all problems have some instances reaching the lower bound,  Note that some QAOA instances have their in-constraint probability below the lower bound, simply because the optimizer fails to find parameters that can satisfy this constraint.

The bottom plots show the fraction of instances that have the optimal solution as the most sampled feasible solution at the end of the optimization. For many instances of the graph partition and the portfolio optimization problem, VQE with in-constraint energy is able to converge to the optimal solution. The other three problems, despite approximation ratios also being high, have fewer instances converging to the optimal solution. This is because the latter problems have solutions that are very close to optimal since the graph weights are only slightly perturbed from $1$ to create a unique optimal solution. These nearly optimal solutions have near-maximum approximation ratios, which distract the in-constraint-energy-oriented optimization. Portfolio optimization naturally does not have this property, and the social network model we used for graph partition tends to produce a distinguished optimal solution. QAOA shows a similar trend, and for all settings, our method leads to a higher probability of converging to the optimal solution compared to the penalty method.

Figure \ref{fig:error bars} compares the quartiles of the metrics over iterations. We observe that the in-constraint energy approach converges extremely fast to a high approximation ratio, while the penalty method improves very slowly. This result is again because the latter fails to balance between the problem's objective and the penalty term and prioritizes optimizing the in-constraint probability. 
As a negative impact, QAOA with the penalty method even slightly lowers the approximation ratio during the optimization process, which verifies the optimizer trace in Figure \ref{fig:grid_search}.
For in-constraint VQE, some instances are able to find the optimal solution and quickly converge to it, but over half of the instances converge to nearly optimal solutions that also have high approximation ratios. For QAOA, both methods show a steady but low fraction of optimal solution in feasible samples.

Figure~\ref{fig:vary} shows the behavior of both methods with different numbers of qubits and QAOA depths. We observe that in-constraint energy keeps being effective for VQE when the number of qubits increases and the penalty method becomes unsuccessful, whereas the performance gain of QAOA with in-constraint energy does not scale well with the number of qubits. The depth of the QAOA has a positive effect on its performance for the first three layers but does not improve afterward.

\section{Discussion}\label{s5}
We propose a new approach for solving constrained combinatorial optimization problems with variational quantum algorithms, using in-constraint energy as the objective. We empirically verify our method's effectiveness under different problem settings and demonstrate significant improvements in the quality of solution samples obtained over the traditional penalty method. We are among the very few studies that consider changing the objective in VQAs, and, to the best of our knowledge, the first that considers adding a constraint to the optimizer to regulate the optimization direction. 

Achieving a commercially relevant quantum advantage in optimization requires tackling the kinds of constrained problems that arise in industrial settings. Since full fault tolerance is believed to not be achievable at practical scales in the very near term, algorithms with low resource requirements provide a promising avenue for evaluating the power of quantum computers to solve optimization problems. By reducing the resource requirements needed to tackle constrained optimization problems, this work brings practical applications of quantum computers one step closer.

\section*{Acknowledgments}
This material is based upon work supported by Q-NEXT, one of the U.S. Department of Energy Office of Science (DOE-SC) National Quantum Information Science Research Centers and the DOE-SC Office of Advanced Scientific Computing Research FAR-QC project under contract number DE-AC02-06CH11357.

\bibliographystyle{myIEEEtran}
\bibliography{ref}

\section*{Disclaimer}

This paper was prepared for information purposes with contributions from the Global Technology Applied Research group of JPMorgan Chase. This paper is not a product of the Research Department of JPMorgan Chase. or its affiliates. Neither JPMorgan Chase nor any of its affiliates make any explicit or implied representation or warranty and none of them accept any liability in connection with this paper, including, but not limited to, the completeness, accuracy, reliability of information contained herein and the potential legal, compliance, tax or accounting effects thereof. This document is not intended as investment research or investment advice, or a recommendation, offer or solicitation for the purchase or sale of any security, financial instrument, financial product or service, or to be used in any way for evaluating the merits of participating in any transaction.

\vfill
\framebox{\parbox{.90\linewidth}{\scriptsize The submitted manuscript has been created by
        UChicago Argonne, LLC, Operator of Argonne National Laboratory (``Argonne'').
        Argonne, a U.S.\ Department of Energy Office of Science laboratory, is operated
        under Contract No.\ DE-AC02-06CH11357.  The U.S.\ Government retains for itself,
        and others acting on its behalf, a paid-up nonexclusive, irrevocable worldwide
        license in said article to reproduce, prepare derivative works, distribute
        copies to the public, and perform publicly and display publicly, by or on
        behalf of the Government.  The Department of Energy will provide public access
        to these results of federally sponsored research in accordance with the DOE
        Public Access Plan \url{http://energy.gov/downloads/doe-public-access-plan}.}}
\end{document}